\documentclass[twocolumn,superscriptaddress,amsmath,amssymb,aps,pra]{revtex4-1}
\usepackage{amsmath,amsthm,amssymb}
\usepackage{latexsym}
\usepackage{amscd,graphicx}
\usepackage[titletoc]{appendix}
\usepackage{epsfig}
\usepackage{epstopdf}
\usepackage[normalem]{ulem}
\usepackage[dvipsnames]{xcolor}
\usepackage[colorlinks=true,linkcolor=red,citecolor=blue]{hyperref}

\begin{document}

\title{Strong Long-Range Spin-Spin Coupling via a Kerr Magnon Interface}

\author{Wei Xiong}

\affiliation{Department of Physics, Wenzhou University, Zhejiang 325035, China}
\affiliation{Interdisciplinary Center of Quantum Information and Zhejiang Province Key Laboratory of Quantum Technology and Device, Department of Physics and State Key Laboratory of Modern Optical Instrumentation, Zhejiang University, Hangzhou 310027, China}

\author{Miao Tian}
\affiliation{Department of Physics, Wenzhou University, Zhejiang 325035, China}

\author{Guo-Qiang Zhang}
\affiliation{Department of Physics, Hangzhou Normal University, Hangzhou 311121, China}
\affiliation{Interdisciplinary Center of Quantum Information and Zhejiang Province Key Laboratory of Quantum Technology and Device, Department of Physics and State Key Laboratory of Modern Optical Instrumentation, Zhejiang University, Hangzhou 310027, China}

\author{J. Q. You}
\altaffiliation{jqyou@zju.edu.cn}
\affiliation{Interdisciplinary Center of Quantum Information and Zhejiang Province Key Laboratory of Quantum Technology and Device, Department of Physics and State Key Laboratory of Modern Optical Instrumentation, Zhejiang University, Hangzhou 310027, China}

\date{\today }

\begin{abstract}
Strong long-range coupling between distant spins is crucial for spin-based quantum information processing. However, achieving such a strong spin-spin coupling remains challenging. Here we propose to realize a strong coupling between two distant spins via the Kerr effect of magnons in a yttrium-iron-garnet nanosphere. By applying a microwave field on this nanosphere, the Kerr effect of magnons can induce the magnon squeezing, so that the coupling between the spin and the squeezed magnons can be exponentially enhanced. This in turn allows the spin-magnon distance increased from nano- to micrometer scale. By considering the virtual excitation of the squeezd magnons in the dispersive regime, strong spin-spin coupling mediated by the squeezed magnons can be achieved, and a remote quantum-state transfer, as well as the nonlocal two-qubit iSWAP gate with high fidelity becomes implementable. Our approach offers a feasible scheme to perform quantum information processing among distant spins.
\end{abstract}


\maketitle

\section{introduction}
Nitrogen-vacancy (NV) center spins in diamond~\cite{hollenberg-2013}, with high controllability~\cite{hollenberg-2013} and long coherence time~\cite{gill-2013,balasubramian-2009}, can be harnessed as a key element in solid-state quantum information networks. {Based on NV center spins, sensing of magnetic fields~\cite{Barry-2020,Shi-2015}, electric fields~\cite{Doherty-2012}, local strain~\cite{Barson-2017}, temperature~\cite{Kucsko-2013}, and quantum information processing and computation~\cite{Ladd-2010} have been studied. However,} it is still a challenge to directly achieve a strong long-distance spin-spin coupling to date, because the usual direct spin-spin coupling is weak and limited by the spatial separation~\cite{kubo2010,marcos2010,zhu2011}. Thus, seeking a good quantum interface to realize a strong indirect spin-spin coupling is significantly important, owing to the no limitation of the indirect coupling by the spin-spin separation.

{At present, lots of quantum systems such as superconducting circuits~\cite{xiong-2018,twamley2010}, mechanical resonators~\cite{Li-2016,Rabl-2010}, cavity optomechanics~\cite{xiong-2021,Chen-2021,Aspelmeyer-2014}, magnons (i.e., the spin-wave excitations or the quanta of spin-wave in magnetic materials)~\cite{Awschalom-2021,Rameshti-2021,wang-2020,quirion-2019,ZhangDK-2015,harder-2018,wang-2019,rao-2021,WangY-2021,xiake1,xiake2,yanpeng2} and non-Hermitian systems~\cite{Xiong1-2021,yanpeng1,Xiong2-2022} have been proposed to play the role of quantum interfaces. Among these,} magnons~\cite{Rameshti-2021,wang-2020,yanpeng2}   in a yttrium-iron-garnet (YIG) sphere, with small mode volume and high spin density, have shown great potential in mediating spin-spin coupling~\cite{neuman-2020,neuman1-2021, neuman2-2021,trifunovic-2013,Fukami-2021,Skogvoll-2021}. For instance, by reducing the size of the YIG sphere from millimeter to nanoscale~\cite{neuman-2020,neuman1-2021, neuman2-2021}, {so that the YIG nanosphere can be regarded as a nanomagnonic cavity, where} the magnetic field in the near-infrared range of the electromagnetic spectrum is concentrated~\cite{Etxarri-2011,Schmidt-2012} for enhancing the local magnetic environment of the spin. {Thus, the strong spin-magnon and spin-spin couplings~\cite{neuman-2020,neuman1-2021, neuman2-2021} can be achieved, and the YIG nanosphere has been proposed to act as the quantum interface for realizing strong spin-photon coupling in a microwave cavity~\cite{hei-2021}}. Also,  magnons in a bulk material~\cite{trifunovic-2013,Fukami-2021} and thin ferromagnet film~\cite{Skogvoll-2021} are used to coherently couple remote spins. However, spins are required to be placed close to or directly deposited on samples for achieving strong spin-magnon interaction~\cite{neuman-2020,neuman1-2021, neuman2-2021,trifunovic-2013,Fukami-2021,Skogvoll-2021}. This imposes difficulties on the manipulation of single spin qubits.

To go beyond the limitations in previous studies~\cite{neuman-2020,neuman1-2021, neuman2-2021,trifunovic-2013,Fukami-2021}, we propose an alternative approach to realize strong long-range spin-magnon and spin-spin couplings by using the Kerr magnons (i.e., the magnons with Kerr effect) in a YIG nanosphere, where spins, located at a micrometer distance from the surface of the YIG nanosphere, are directly but weakly coupled to the magnons in the nanosphere driven by a microwave field.  Experimentally, strong and tunable Kerr nonlinearity of the magnons, originating from the magnetocrystalline anisotropy in the YIG sphere~\cite{zhang-2019}, has been  demonstrated~\cite{wang-2016}. This nonlinear effect in magnonics has be used to study the bi- and multi-stabilities~\cite{wang-2018,nair-2020,shen-2021}, as well as the nonreciprocity~\cite{kong-2019}, quantum entanglement~\cite{zhangz-2019} and quantum phase transition~\cite{zhang-2021}. The microwave field acting on the Kerr magnons can give rise to a strong magnon squeezing effect. With this squeezing effect, the spin-magnon coupling can be exponentially enhanced. Thus, by using suitable parameters, the enhanced spin-magnon coupling can enter the strong and even ultrastrong coupling regimes. Such a strong coupling allows the quantum state to transfer between the spin and the squeezed magnons. Furthermore, we study the enhanced spin-magnon system in the dispersive regime, so as to obtain a strong long-range spin-spin coupling by adiabatically eliminating the degrees of freedom of the squeezed magnons. This spin-spin coupling can be used to exchange quantum states of remote spins.

In contrast to Refs.~\cite{neuman-2020,neuman1-2021, neuman2-2021,trifunovic-2013,Fukami-2021}, a tunable Kerr effect is introduced in our approach to obtain a controllable long-distance spin-magnon coupling. Also, it is different from Ref.~\cite{Skogvoll-2021} that the YIG nanosphere is used here instead of a thin ferromagnet film, which can offer a stronger Kerr effect~\cite{zhang-2019}. Moreover, a driving field is applied on the YIG nanosphere to enhance the two-magnon process via linearizing the magnon Kerr effect, so as to achieve a strong magnon squeezing needed in the approach. {More importantly, the spin is remotely placed away from the YIG nanosphere in our proposal. This leads to control of the single spin qubit feasible. But in Ref.~\cite{Skogvoll-2021}, the {\it intrinsic weak} Kerr effect is employed, which actually gives rise to a small squeezing parameter and thus coupling amplification is limited. So in principle the spin qubit is also required to be close to the YIG thin film for obtaining the large magnon-spin coupling.}

\begin{figure}
	\center
	\includegraphics[scale=0.35]{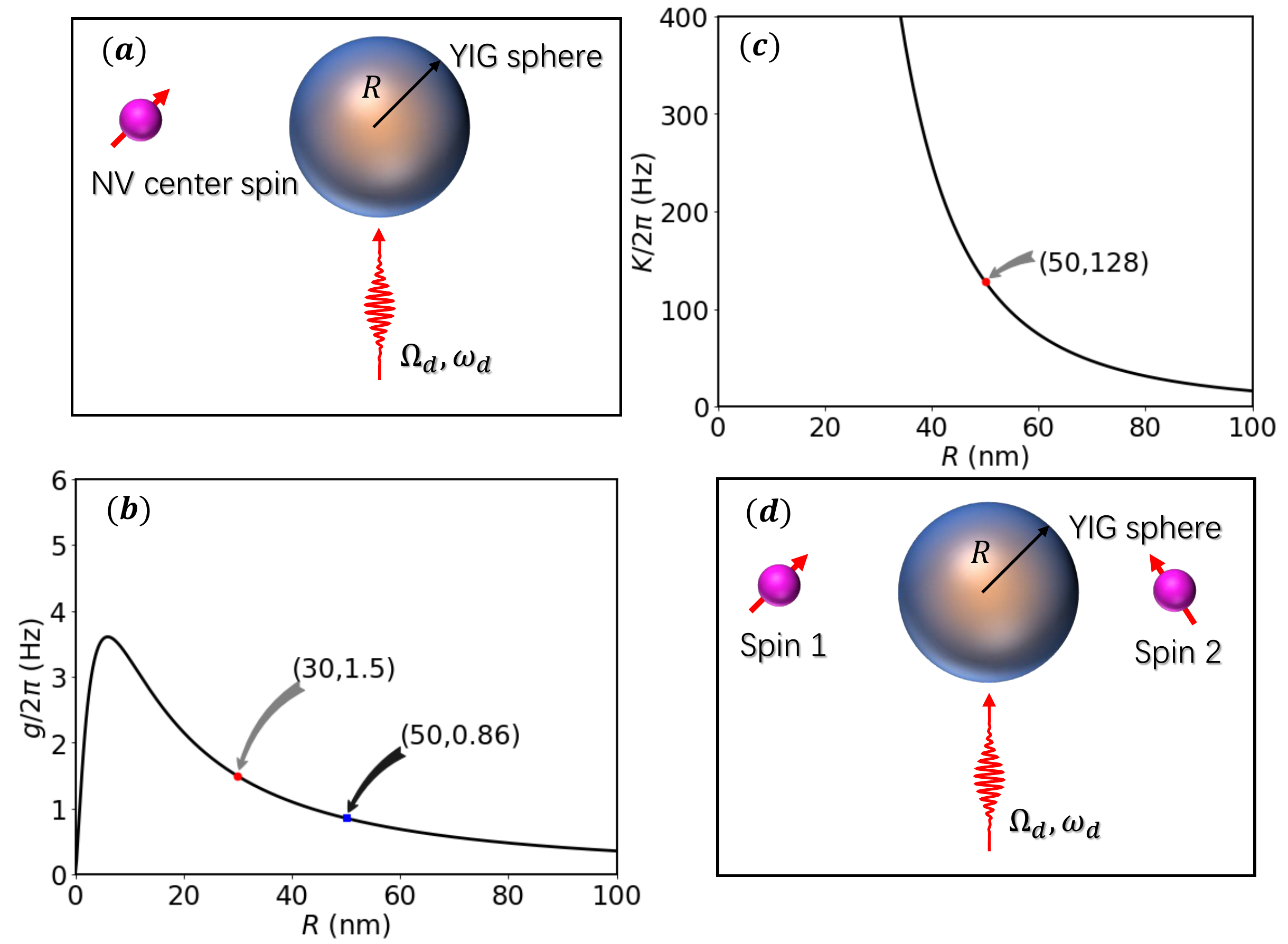}
	\caption{(a) Schematic of the proposed hybrid system consisting of a NV-center spin weakly coupled to the Kerr magnons in a YIG sphere with the {nanometer} radius $R$. The NV-center spin is located $d$ distance from the sphere surface, and the magnons are driven by a microwave field with frequency $\omega_d$ and amplitude $\Omega_d$. (b) The coefficient of the magnon Kerr nonlinearity $K/2\pi$ versus the radius $R$ of the YIG nanosphere. (c) The spin-magnon coupling versus $R$. (d) Schematic of two NV-center spins weakly coupled to the Kerr magnons driven by a microwave field.}\label{fig1}
\end{figure}

\section{ The model Hamiltonian} We consider a hybrid quantum system consisting of a single NV center in diamond as a spin qubit with transition frequency $\omega_q$, weakly coupled to the Kerr magnons in a YIG sphere with the nanometer radius $R$ [see Fig.~\ref{fig1}(a)]. The Kerr nonlinearity of the magnons stems from the magnetocrystalline anisotropy~\cite{zhang-2019,wang-2016}, and the spin qubit is placed at a distance $d$ from the surface of the YIG sphere. The Hamiltonian of the hybrid system under the rotating-wave approximation (RWA) is (setting $\hbar=1$)
\begin{align}
H_{\rm  NL}=\frac{1}{2}\omega_q\sigma_z+H_K+g(\sigma_+ m+m^\dagger \sigma_-),\label{eq1}
\end{align}
where $\sigma_\pm$ are the lowering and rising operators of the spin qubit and $m~(m^\dag)$ is the annihilation (creation) operator of the magnons in the YIG sphere. The Kerr Hamiltonian $H_K=\omega_m m^\dag m-(K/2) m^\dag m^\dag m m$, with the angular frequency $\omega_m=\gamma B_0+2\mu_0K_{\rm an}\gamma^2/(M^2V_m^2)-2\mu_0\rho_s sK_{\rm an}\gamma^2/M^2$ and the Kerr coefficient $K/\hbar=2\mu_0K_{\rm an}\gamma^2/(M^2V_m^2)$, represents the interaction among magnons and provides the anharmonicity of the magnons, where $\gamma/2\pi=g_e\mu_B/\hbar$ is the gyromagnetic ration with the $g$-factor $g_e$ and the Bohr magneton $\mu_B$, $\rho_s=2.1\times10^{-22}~{\rm cm}^{-3}$ is the spin density of the YIG sphere, $\mu_0$ is the vacuum permeability, $K_{\rm an}$ is the first-order anisotropy constant of the YIG sphere, $B_0$ is the amplitude of a bias magnetic field along $z$-direction, $M$ is the saturation magnetization, and $V_m$ is the volume of the YIG sphere. Here $g/(2\pi)=\sqrt{|\gamma|MR^{3}/(24\pi\hbar)}g_e\mu_0\mu_B /(d+R)^3$ is the coupling strength between the spin qubit and the magnons~\cite{hei-2021}, which first increases to its maximum and then decreases with the size of the YIG sphere [see Fig.~\ref{fig1}(b) for $d=6$ nm]. When $R\sim30$ nm, $g/2\pi\sim 1.5$ KHz, and $R\sim50$ nm gives $g/2\pi\sim 0.86$ KHz~\cite{neuman-2020,hei-2021}. In addition, the spin-magnon coupling $g$ can sharply decrease when increasing the spin-magnon separation $d$. Because the Kerr coefficient is inversely proportional to the volume of the YIG sphere, i.e., $K\propto V_m^{-1}$, the Kerr effect can become significantly important for a YIG nanosphere [see Fig.~\ref{fig1}(c)]. For example, when $R\sim50$ nm, $K/2\pi\sim128$ Hz, but $K/2\pi=0.05$ nHz for $R\sim0.5$ mm (the usual size of the YIG sphere used in various previous experiments). Obviously, $K$ is much smaller in the latter case. {Note that the magnons in YIG nanosphere can interact with the vibration mode supporting phonons via magnetostrictive interaction. Such the magnon-phonon coupling gives to magnon Kerr nonlinearity, leading to magnon-spin coupling (de)amplification by tuning $K$~\cite{sm}}.

With the strong Kerr effect in the YIG sphere, a strong long-distance spin-magnon coupling becomes achievable. To realize this, a microwave field (MWF) with  amplitude $\Omega_d$  and frequency $\omega_d$ is applied to the magnons, which can be characterized by the Hamiltonian $H_D=\Omega_d (m^\dag e^{-i\omega_dt}+m e^{i\omega_dt})$. Thus, the total Hamiltonian of the hybrid system is $H_{\rm total}=H_{\rm NL}+H_D$. By rewritting the magnon operator as the expectation value plus its fluctuation, i.e., $m\rightarrow \langle m\rangle+m$, the Kerr Hamiltonian $H_K$ in Eq.~(\ref{eq1}) can be linearized by neglecting the higher-order fluctuation terms, as guaranteed by a strong driving MWF~\cite{sm}. In the rotating frame with respect to $\omega_d$, Eq.~(\ref{eq1}) reduces to~\cite{sm}
\begin{align}
H_{L}=\frac{1}{2}\Delta_q\sigma_z+H_{\mathcal{K}}+g(\sigma_+ m+m^{\dag}\sigma_-),\label{eq2}
\end{align}
with
\begin{align}
	H_\mathcal{K}={\Delta}_m m^\dag m-\frac{1}{2}\mathcal{K}(m^2 +m^{{\dag}2}),\label{eq2'}
\end{align}
where $\Delta_{m}=\omega_{m}+2K N_m-\omega_d$, with $N_m=|\langle m\rangle|^2$, is the magnon-number-dependent frquency detuning, $\Delta_q=\omega_q-\omega_d$ is the spin-qubit frequency detuning, and
$\mathcal{K}=K\langle m\rangle^2 $ is the enhanced coefficient of the two-magnon process, induced by the linearization of the Kerr magnons. As $K$ can be either positive or negative by adjusting the crystallographic axis $[100]$ or $[110]$ of the YIG sphere along the bias magnetic field~\cite{wang-2016}, we can have $\mathcal{K}>0$ or $\mathcal{K}<0$. {In addition, $\mathcal{K}$ is much larger and more tunable than the one obtained in Ref.~\cite{Skogvoll-2021}. This is because the Kerr coefficient $K$ can be larger here for a YIG nanosphere, and the value of $\langle m\rangle$ is controlled by the applied driving field.} The linearized Kerr Hamiltonian $H_\mathcal{K}$ in Eq.~(\ref{eq2'}) describes the two-magnon process, which can give rise to the magnon squeezing needed in the approach.

\section{Strong and ultrastrong long-range spin-magnon couplings}
Long-range light-matter interaction is important for quantum information processing. Previous proposals~\cite{neuman-2020,neuman1-2021, neuman2-2021} show that a strong spin-magnon coupling can only be attained when the spin is extremely close to the surface of the YIG sphere, i.e., $d\sim 10$~nm. In our approach, to realize the strong long-range spin-magnon interaction, we operate the system in the squeezed-magnon frame via the Bogoliubov transform $m= m_s\cosh r_m+m_s^\dag \sinh r_m$. Substituting this expression into Eq.~(\ref{eq2}) and setting the squeezing parameter $r_m=(1/4)\ln(\Delta_m+\mathcal{K})/(\Delta_m-\mathcal{K})$, we can reduce Eq.~(\ref{eq2}) to~\cite{sm}
\begin{align}
	H_{\rm S}=&\frac{1}{2}\Delta_q\sigma_z+\Delta_s m_s^\dag m_s+G (m_s^\dag+m_s)(\sigma_++\sigma_-),\label{eq3}
\end{align}
which is the standard quantum Rabi model with a controllable frequency $\Delta_s=\sqrt{\Delta_m^2-\mathcal{K}^2}$ and an exponentially enhanced coupling, $G=ge^{r_m}/2$, between the spin and the squeezed magnons.

\begin{figure}
	\center
	\includegraphics[scale=0.3]{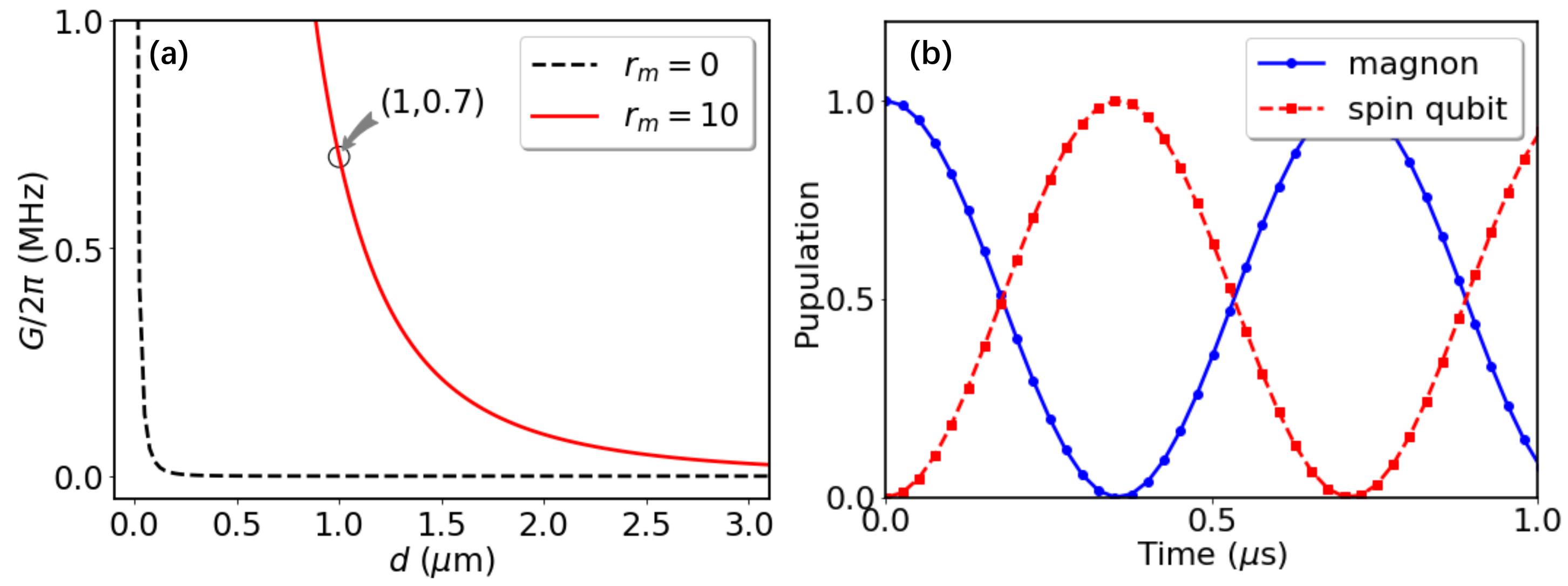}
	\caption{(a) The spin-magnon coupling strength versus the distance between the spin qubit and the surface of the YIG nanosphere with the squeezing parameter $r_m=0$ and $r_m=10$. (b) Populations of the magnons and spin qubit versus the evolution time with $G=4$ MHz and $\Delta_q=\Delta_s=10G$.}\label{fig2}
\end{figure}

Since the parameters $\Delta_m$ and $\mathcal{K}$ are both tunable, the squeezing parameter $r_m$ can be very large when the instability threshold approaches, i.e., $\Delta_m=\mathcal{K}$. This directly yields the enhanced spin-magnon coupling several orders of magnitude larger than the original one. Such an enhancement allows the distance between the spin qubit and the magnons from the nanometer to micrometer scale. To address this, the exponentially enhanced spin-magnon coupling versus the spin-magnon separation is plotted in Fig.~\ref{fig2}(a). Without the magnon squeezing ($r_m=0$), the spin-magnon coupling sharply decreases to zero when $d$ is increased to dozens of nanometer. With the magnon squeezing (e.g., $r_m=10$), we find that the enchanced coupling $G$ can be several megahertz for $d$ approaching to $1~\mu$m, i.e., $G\sim 4$~MHz for $d=1~\mu$m. Compared to Refs.~\cite{neuman-2020,neuman1-2021, neuman2-2021}, the spin-magnon separation is increased {\it three} orders of magnitude when maintaining the same large spin-magnon coupling. {In addition, the instability threshold also gives a boundary for the amplitude of the MWF~\cite{sm}}.

For the NV center spin and the magnons, the typical decay rates are $\gamma_q\sim1$ KHz and $\kappa_m\sim1$ MHz, respectively. Thus, the enhanced spin-magnon system can be in the {\it strong} coupling regime, as justified by the Rabi oscillation in Fig.~\ref{fig2}(b), where the populations of the magnons and the spin qubit versus the evolution time are plotted at $\Delta_q=\Delta_s=10G$. Obviously, the quantum state exchange between the magnons and the spin qubit can occur, where the squeezed magnons are initially excited and the spin qubit is in the ground state.  Here both $\Delta_s$ and $\Delta_q$ are controllable via tuning the frequency of the MWF, so $G$ can be comparable to or exceed $\Delta_{q(s)}$. This suggests that the enhanced spin-magnon system can be even in the {\it ultrastrong} coupling regime with the assistance of the MWF. Moreover, the unwanted terms in Eq.~(\ref{eq3}) is greatly suppressed~\cite{sm} due to the large squeezing parameters $r_m$. This can be achieved in our proposal by tuning the crystallographic axis $[100]$ of the YIG sphere along the bias magnetic field, i.e., $K>0$.

\begin{figure}
	\center
	\includegraphics[scale=0.38]{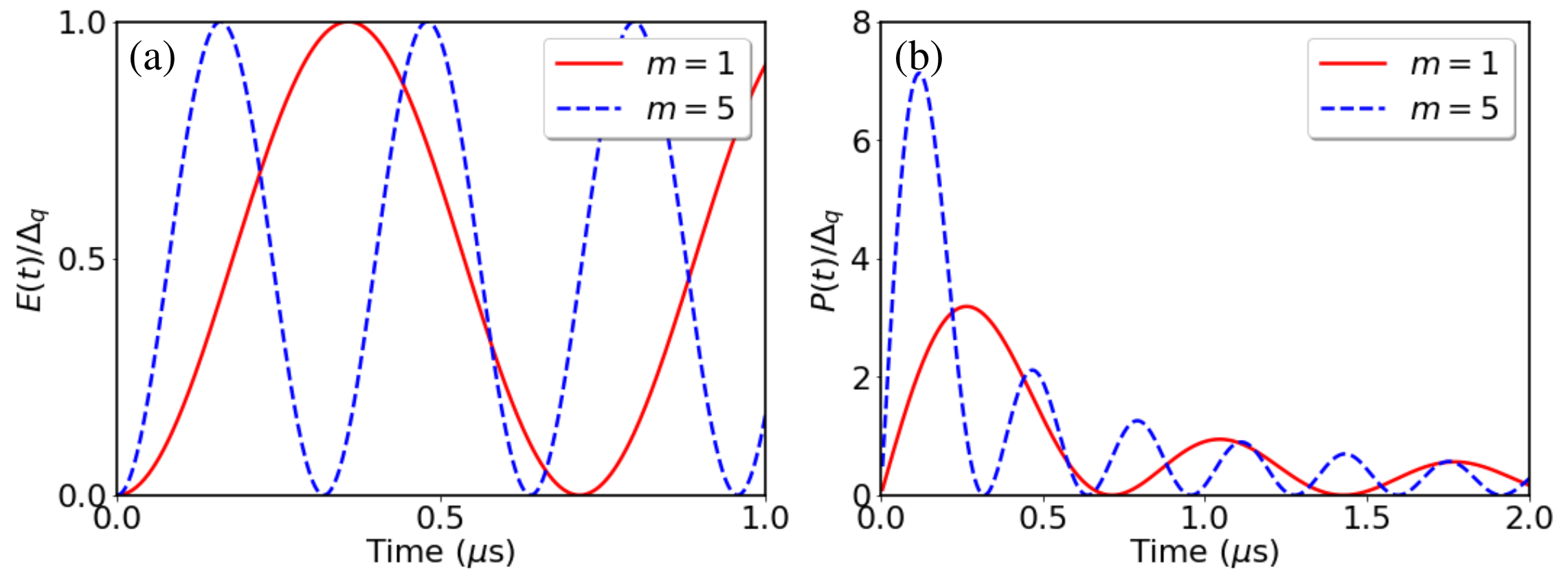}
	\caption{(a) The energy stored in a single spin battery and (b) the corresponding power versus the evolution time for different magnon excitations $m=1,~5$. Here $\Delta_s=\Delta_q=10 G$ and $G=4$ MHz.}\label{fig4}
\end{figure}

We can use the strong long-distance spin-magnon coupling in the proposed hybrid system to implement a quantum battery~\cite{binder-2015,campaioli-2017,ferraro-2018,jing-2021}. In this thermal device, the energy stored in the charger acted by the magnons can be transferred to the battery played by the NV center spin. Initially, the charger is in the state $|m\rangle$ with $m$ excitations, and the battery is in the ground state $|g\rangle$. Governed by the Hamiltonian (\ref{eq3}) with the RWA, the energy of the quantum battery and the corresponding power oscillate with the evolution time~[see Fig.~\ref{fig4}(a)]. At a certain time, i.e., $t=\pi/(2G)$, the battery is fully charged and the maximum power is reached~[see Figs.~\ref{fig4}(a) and (b)].  More importantly, we find that the excitations in the magnon mode can speed up the charging process by a factor of $\sqrt{m}$ [see Fig.~\ref{fig4}(a)] and the power is hence greatly enhanced by $\sqrt{m}$. This is due to the simultaneous charging of the battery by $m$ chargers with energy $\hbar\Delta_s$.

\section{Long-range spin-spin coupling mediated by the squeezed magnons}
Now we consider the case that two spins with the same effective frequency $\Delta_q$, placed at $d\sim 1$~$\mu$m distance from the surface of the YIG nanosphere, are coupled to the squeezed magnons [see Fig.~\ref{fig1}(d)]. As $d\gg R\sim 30$~nm, the size of the YIG nanosphere can be ignored and thus the separation between the two spins is $\sim 2$~$\mu$m. When the condition $G\ll \Delta_+\equiv\Delta_s+\Delta_q$ is satisfied, the RWA is allowed for Eq.~(\ref{eq3}) and hence the counter-rotating terms can be neglected. Now, we have
\begin{align}
	H_{R}=\Delta_s m_s^\dag m_s+\sum\limits_{i=1}^{2}\left[\frac{1}{2}\Delta_q\sigma_z^{(i)}+G(m_s^\dag\sigma_-^{(i)}+{\rm H.c.})\right],\label{eq5}
\end{align}
which is a typical Tavis-Cummings model. We further consider the spin-magnon system in the {\it dispersive} regime, i.e., $G\ll|\Delta_-\equiv\Delta_s-\Delta_q|$, which allows to obtain a strong effective coupling between the two remote spins. This can be achieved by adiabatically eliminating the degrees of freedom of the squeezed magnons via Fr\"{o}hlich-Nakajima transformation~\cite{frohlich-1953,nakajima-1953}.

\begin{figure}
	\center
	\includegraphics[scale=0.39]{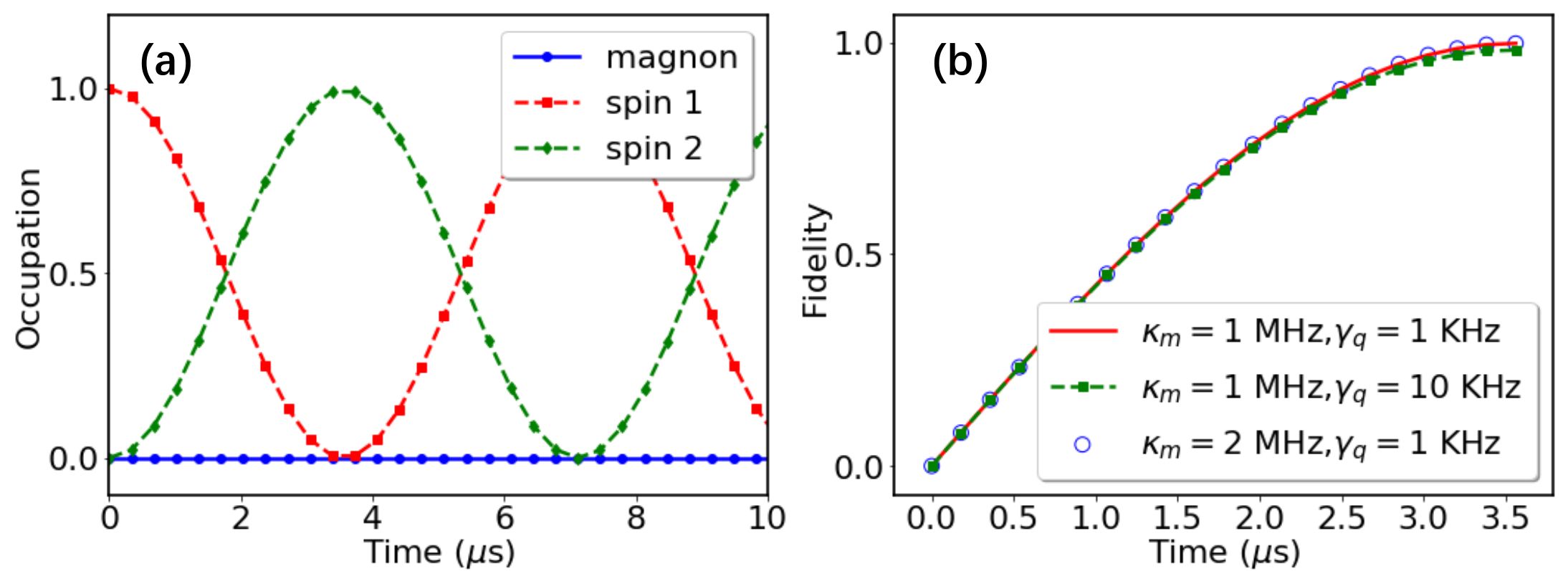}
	\caption{(a) Occupations of the spin 1, spin 2 and magnons versus the evolution time by {numerically} solving Eq.~(\ref{eq6}) with the dissipations included. (b) The fidelity of the two-qubit iSWAP gate versus the evolution time by considering the dissipations. Here $\kappa_m=1$ MHz, $\gamma_q=1$ KHz, and $G_{\rm eff}=70$ KHz.}\label{fig3}
\end{figure}

The effective spin-spin coupling Hamiltonian is obtained as~\cite{sm}
\begin{align}
H_{\rm eff}=\frac{1}{2}\omega_{\rm eff}(\sigma_{z}^{(1)}+\sigma_{z}^{(2)})+G_{\rm eff}(\sigma_{+}^{(1)}\sigma_{-}^{(2)}+{\rm H.c.}),\label{eq6}
\end{align}
where $\omega_{\rm eff}=(1+2\langle m_s^\dag m_s\rangle)\Delta_q^2/\Delta_-$ is the magnon-number-dependent frequency of each spin induced by the spin-magnon dispersive coupling, and $G_{\rm eff}=G^2/\Delta_-$ is the effective spin-spin coupling strength. To estimate this strength, we take $G\sim2\pi\times0.7$ MHz for $d\sim1~\mu$m and $\Delta_-\sim 10G$, so $G_{\rm eff}/2\pi\sim 70$ KHz $\gg \gamma_q\sim 1$ KHz. This strong long-range spin-spin coupling enables the quantum state transfer between two spins mediated by the magnons, and the two-qubit iSWAP gate becomes achievable, i.e.,
\begin{align}
&|g_1\rangle|g_2\rangle\rightarrow|g_1\rangle|g_2\rangle, \nonumber\\
&|g_1\rangle|e_2\rangle\rightarrow -i|e_1\rangle|g_2\rangle,\nonumber\\
&|e_1\rangle|g_2\rangle\rightarrow -i|g_1\rangle|e_2\rangle,\\
&|e_1\rangle|e_2\rangle\rightarrow |e_1\rangle|e_2\rangle, \nonumber
\end{align}
where $|g\rangle$ and $|e\rangle$ denote the ground and excited states of the spin qubit, respectively.

The relevant quantum dynamics is governed by
\begin{align}
\dot {\rho}=&i[\rho, H_{\rm eff}]+\kappa_{m}\mathcal{D}[m_s]\rho+\gamma_q\mathcal{D}[\sigma_-]\rho,\label{eq4}
\end{align}
where $\mathcal{D}[x]=x\rho x^\dag-(x^\dag x\rho+\rho x^\dag x)/2$, with $x=m_s$,$\sigma_-^{(1)}$ and $\sigma_-^{(2)}$, while $\gamma_q$ denotes the transversal relaxation rate
of the NV-center spin. {Here we only consider the case of the magnon mode in the zero temperature environment ($T=0$). The case of the finite temperature ($T\neq0$) is also discussed~\cite{sm}}. Experimentally, the longitudinal relaxation rate is much smaller than $\gamma_q$~\cite{angerer-2017}, so we can ignore its effect on system's dynamics.
In Fig.~\ref{fig3}, we plot the occupation probability and gate fidelity versus the evolution time, where we initially prepare spin 1 in the excited state, spin 2 in the ground state, and the magnons in the ground state. From Fig.~\ref{fig3}(a) we see that in the presence of dissipations ($\kappa_m=1$ MHz and $\gamma_q=1$ KHz), quantum state from spin 1 (2) can be transferred to spin 2 (1), while the magnons are kept in the initial state, due to the dispersive spin-magnon coupling.  Thus, the fidelity of the two-qubit iSWAP gate is robust against the decay rate of the magnons [see Fig.~\ref{fig3}(b)]. Also, {we find that the fidelity of this two-qubit iSWAP gate is slightly affected by the decay rate of the spin qubit}, owing to the achieved strong spin-spin coupling (i.e., $G_{\rm eff}\gg \gamma_q$) [see Fig.~\ref{fig3}(b)]. {Besides, the effect of the Gilbert damping parameter on the fidelity of the two-qubit iSWAP gate is stuided~\cite{sm}}. 

\section{Discussion and Conclusion}

{In our proposal, the magnetic moment in YIG comes from ${\rm Fe}^{3+}$ ions in the ground state
$^6S_{5/2}$. Since the ground state has no orbital angular momentum, there should
be no spin–orbit coupling and hence no anisotropy~\cite{Stancil-2009}. Experimentally, however,
a small anisotropy energy contribution is present. This is probably due to
small spin–orbit interactions among the electronic substates neglected in the
Russell–Saunders coupling scheme\cite{Vleck-1934}. In other words, the possibly existed spin-orbit coupling may give rise to the magnetocrystalline anisotropy~\cite{Soumyanarayanan-2016}, which leads to the Kerr nonlinearity demonstrated experimentally~\cite{wang-2016}. In our proposal, the Kerr nonlinearity as a key element is employed to amplify the magnon-spin coupling. With the smaller YIG sphere, the bigger Kerr coefficient is obtained, which gives larger squeezing parameter and stronger spin-magnon coupling. So we here only consider the Kerr effect (or magnetocrystalline anisotropy) instead of the possibly existed spin-orbit coupling.} 

{At the present stage, the YIG nanosphere has not been fabricated experimentally, although the millimeter-sized YIG spheres have been widely used to demonstrate abundant phenomena in quantum physics. With the nanotechnology development, the nanomagnets have been realized~\cite{Wang-2021,Hou-2019}. Therefore, generating YIG nanospheres expermentlly is promising in future. Our scheme is also dependent on the Kerr nonlinearity of the YIG sphere. Fortunately, this nonlinear effect has been demonstrated very recently~\cite{wang-2016}. Based on these, the design and control of magnetic nanostructures for realizing strong spin-magnon coupling and magnon-mediated spin-spin coupling remains an open question for future studies in the field of cavity nanomagnonics.}

In summary, we have proposed a feasible scheme to achieve a strong long-range spin-spin coupling via the Kerr mangons in a YIG nanosphere. Assisted with the drive field, the Kerr nonlinearity of the magnons can be converted to the magnon squeezing effect, which exponentially enhances the spin-magnon coupling in the squeezed-magnon frame with the experimentally available parameters. This strong coupling allows the separation between the spins and the YIG nanosphere increased from nano- to micrometer scale. In the dispersive regime, the virtual squeezed magnons can be adiabatically eliminated to yield a strong long-range spin-spin coupling. With these strong spin-magnon and spin-spin couplings, some quantum information processing tasks such as the remote quantum-state transfer and high-fidelity two-qubit iSWAP gate are implementable. Our proposal can provide a potential platform to realize quantum information processing among remote spins by using a hybrid system consisting of the Kerr magnons coupled to the spin qubits.

\begin{acknowledgments}
This work is supported by the National Natural Science Foundation of China (Grants No.~11934010, No.~U1801661, and No.~U21A20436),  the key program of the Natural Science Foundation of Anhui (Grant No.~KJ2021A1301).
\end{acknowledgments}

\end{document}